\documentclass{article}
\usepackage{amssymb}
\usepackage{amsfonts}
\usepackage{amsmath}
\usepackage{graphicx}
\usepackage{appendix}

\newtheorem{theorem}{Theorem}
\newtheorem{acknowledgement}[theorem]{Acknowledgement}

\begin{document}

\title{Garrett approximation revisited}
\author{Victor Barsan \\
IFIN-HH, 30 Reactorului\\
and \\
UNESCO\ Chair of HHF, 407 Atomistilor\\
077125 Magurele, Romania}
\maketitle

\begin{abstract}
Three variants of the Garrett approximation are studied and their accuracy
is analyzed, for symmetric and asymmetric square wells. Quite surprisingly,
the simplest variants are also the most accurate. The applications to
quantum wells, quantum dots and capillary neutron guides are briefly
discussed.
\end{abstract}

\section{Introduction}

Surprisingly or not, one of the most elementary problems of quantum
mechanics - a particle in a symmetric square well - is still under debate:
if its wave function can be easily expressed in terms of elementary
functions, the bound states energy eigenvalues are given by transcendental
equations, which defy exact solutions. A large number of approximations
was proposed - based on graphical constructions \cite{[Schiff]}, \cite%
{[Bohm]}, \cite{[Pitkanen]}, on mathematical tricks \cite{[Barker]} \cite%
{[dABG]}, \cite{[VB-RD]}, \cite{[VB-RRP]} or physical ideas \cite{[Garrett]}%
. In this paper, we shall pay attention to an approach based on a physical
idea, due to Garrett \cite{[Garrett]}: as the main difference between the
infinite and finite square well is the fact that, in the first case, the
wall is impenetrable, but in the second one, the wave function penetrates
the wall on a certain distance $\delta $, the energy of a bound state $E_{n}$
in a finite well of length $L$ should be satisfactorily approximated by the
energy of the corresponding bound state, $E_{n}^{\left( 0\right) }$, in an
infinite well of length $L+2\delta $.

Garrett's idea is interesting from educational point of view, as it provides
a way of understanding quantum phenomena without solving Schr\"{o}dinger
equation \cite{[VB-EJP]}; it was recently discussed in textbooks \cite%
{[Cahay]}. More than this, it has several applications in the theoretical
description of quantum wells \cite{[Badescu-1997]}, quantum dots \cite%
{[Ind-1999]}, capillary neutron guides \cite{[Roh-2002]} and infrared
photodetectors \cite{[VB-arXiv-2018]}. A more attentive analysis pointed out
that, besides the original approach proposed by Garrett, there are two more
variants of this approximation \cite{[VB-RD]}, \cite{[VB-EJP]} which, in
some cases, may provide more accurate results. The goal of this paper is to
make a detailed (mainly numerical) investigation of the accuracy of each of
these three variants, in the calculation of the energy levels of several
symmetric and asymmetric square wells. The application of Garrett approach
to asymmetric wells is appealing, as it could be used for a simple and quite
precise evaluation of energy levels in stepped wells, so important for
semiconductor heterostructures (see for instance \cite{[Lamberti]}).

The paper has the following structure. The second section is merely a
reformulation of previous results \cite{[Garrett]}, \cite{[VB-RD]}, \cite%
{[VB-EJP]}, i.e. we introduce the three variants of Garrett approximation
and obtain convenient formulas for the calculation of dimensionless wave
vectors of the bound states in the square well. It is the starting point for
the evaluation of errors of each variant, implicitly of determining its
adequacy for a certain bound state. The same scheme is applied to the Barker
approximation. The third section is a comparative analysis of Garrett and
Barker approximations for finite square wells: we find out which variant is
the most appropriate (i. e. the most precise) for a specific case; the most
relevant results are conveniently presented as plots - and tables, included as auxiliary material. In the fourth section, the same
treatment is applied to the simple asymmetric well. In the fifth one, we
discuss the applications of Garrett approach to quantum dots and capillary
neutron guides. The last one is devoted to conclusions.

\section{Garrett's approximation for the bound states energy of finite
square wells}

In order to expose Garrett's approach, let us mention that the energy levels
of a particle of mass $m$\ in an infinite rectangular well of length $L$\ is
given by the well-known formula:

\begin{equation}
E_{n}^{\left( 0\right) }=n^{2}\frac{\pi ^{2}\hbar ^{2}}{2mL^{2}}  \label{1}
\end{equation}

The same particle, moving in a finite square well of depth $V$ and same
length, can propagate in the classically forbidden
region, where its wave function decays exponentially with a characteristic
length $\delta :$

\begin{equation}
\delta =\frac{\hbar }{\sqrt{2m\left( V-E\right) }}  \label{2}
\end{equation}%
whith $E$ - the energy of its bound state. Garrett notices that "the use of
this length to modify the effective width of the infinite well will lead to
a simple iterative approximation for the energy states of the finite well".

In the first iteration, the energy $E_{n}^{\left( 0\right) }$ of the $n-$th
level of the infinite well (\ref{1}) can be introduced into (\ref{2}), to
provide the first approximation for the penetration of the $n-$th wave
function of the finite well into the classically forbidden region:

\begin{equation}
\delta ^{\left( 1\right) }=\frac{\hbar }{\left[ 2m\left( V-E_{n}^{\left(
0\right) }\right) \right] ^{1/2}}  \label{3}
\end{equation}%
So, making in (\ref{1}) the substitution $L\rightarrow L+2\delta ^{\left(
1\right) },$ we shall obtain a first approximation of the $n-$th state of
the finite well:

\begin{equation}
E_{n}^{\left( 1\right) }=n^{2}\frac{\pi ^{2}\hbar ^{2}}{2m\left( L+2\delta
^{\left( 1\right) }\right) ^{2}}  \label{4}
\end{equation}

In the second iteration, we can substitute $E_{n}^{\left( 1\right) }$ into (%
\ref{3})

\begin{equation}
\delta ^{\left( 2\right) }=\frac{\hbar }{\left[ 2m\left( V-E_{n}^{\left(
1\right) }\right) \right] ^{1/2}}=\frac{\hbar \left( L+2\delta ^{\left(
1\right) }\right) }{\left[ 2mV\left( L+2\delta ^{\left( 1\right) }\right)
^{2}-\pi ^{2}\hbar ^{2}n^{2}\right] }  \label{5}
\end{equation}%
and get a second order correction of the penetration length, to be used to
the substitution $L\rightarrow L+2\delta ^{\left( 2\right) }$ in (\ref{1}),
providing a second order approximation of the $n-$th state of the finite
well:%
\begin{equation}
E_{n}^{\left( 2\right) }=n^{2}\frac{\pi ^{2}\hbar ^{2}}{2m\left( L+2\delta
^{\left( 2\right) }\right) ^{2}}  \label{6}
\end{equation}

Defining $P,$ a dimensionless quantity, characterizing both the potential $%
\left( L,\ V\right) $\ and the particle $\left( m\right) ,$ and its inverse $%
1/p$

\begin{equation}
P=\sqrt{2mV}\frac{L}{2\hbar }=\frac{1}{p}  \label{7}
\end{equation}%
and noticing that

\begin{equation}
\frac{E_{n}^{\left( 0\right) }}{V}=\frac{\pi ^{2}n^{2}}{4}p^{2}  \label{8}
\end{equation}%
we can write the penetration lengths in a dimensionless form:

\begin{equation}
\frac{2\delta ^{\left( 1\right) }}{L}=\frac{p}{\sqrt{1-\frac{\pi ^{2}n^{2}}{4%
}p^{2}}}  \label{9}
\end{equation}%
\begin{equation}
\frac{2\delta _{n}^{\left( 2\right) }}{L}=\frac{p}{\left( 1-\frac{\pi
^{2}n^{2}}{4}\frac{p^{2}}{\left( 1+\frac{p}{\left( 1-\frac{\pi ^{2}n^{2}}{4}%
p^{2}\right) ^{1/2}}\right) ^{2}}\right) ^{1/2}}  \label{10}
\end{equation}

Clearly, higher order iterations will produce too cumbersome expressions,
instead of (\ref{9}) and (\ref{10}), but, by the substitutions $\delta ^{\left(
1\right) }\rightarrow \delta ^{\left( l\right) }$ and $\delta ^{\left(
2\right) }\rightarrow \delta ^{\left( l+1\right) }$, the relation (\ref{5})
will be replaced by an equally simple one:

\begin{equation}
\delta ^{\left( l+1\right) }=\frac{\hbar \left( L+2\delta ^{\left( l\right)
}\right) }{\left[ 2mV\left( L+2\delta ^{\left( l\right) }\right) ^{2}-\pi
^{2}\hbar ^{2}n^{2}\right] }  \label{11}
\end{equation}%
where the index $n$ of $\delta _{n}^{\left( l\right) }$ was dropped, in
order to avoid too complicated notations. Taking the limit $l\rightarrow
\infty $ in (\ref{11}) and defining:

\begin{equation}
\lim_{q\rightarrow \infty }\delta ^{\left( q\right) }=\Delta ,\ y=\frac{%
2\Delta }{L}  \label{12}
\end{equation}%
we get for the dimensionless penetration depth $y$:

\begin{equation}
4P^{2}y^{4}+8P^{2}y^{3}+\left( 4P^{2}-\pi ^{2}n^{2}-4\right) y^{2}-8y-4=0
\label{13}
\end{equation}%
or, equivalently, with $P\rightarrow \frac{1}{p}$:

\begin{equation}
y^{4}+2y^{3}+\left( 1-\left( \frac{\pi ^{2}n^{2}}{4}+1\right) p^{2}\right)
y^{2}-2p^{2}y-p^{2}=0  \label{14}
\end{equation}%
It is easy to check that, for deep wells, i.e. for small values of $p,$ in
the first order approximation, $\ y\left( p\right) \simeq p;$ so, the
quartic and the cubic terms in (\ref{14}) can be neglected. Also, for deep
levels, $\left( \frac{\pi ^{2}n^{2}}{4}+1\right) p^{2}\ll 1,$ and (\ref{14})
becomes:

\begin{equation}
y^{2}-2p^{2}y-p^{2}=0  \label{15}
\end{equation}%
with the positive root:

\begin{equation}
y=p\sqrt{p^{2}+1}+p^{2}\simeq p+p^{2},\ p\ll 1,\ n\sim 1  \label{16}
\end{equation}

Let us comment now on the Garrett's iterative approximation. In his original
paper, he uses only two iterations. The result obtained in this way, Eq. (%
\ref{10}) (which was not explicitly written by Garrett) discourages the
attempt of going to higher orders. However, it is easy to apply consistently
Garrett's idea, i.e. to consider an infinite number of iterations, according
to Eqs. (\ref{11}-\ref{14}). In this situation, it would be interesting to
investigate the following aspects:

(*1) The consistent application of Garrett's idea (considering an infinite
number of iterations, which generates a quartic equation, (\ref{14})) gives
better results than Garrett's original two-iteration approach, Eq. (\ref{10}%
) ?

(*2) The simple approximation of the roots of the quartic equation, so
restrictive, independent of the index of energy level $n$, obtained for
large wells and deep levels, (\ref{15}), (\ref{16}), can provide useful
results?

(*3) For practical applications, which one is more convenient: the
"consistent" approximation (\ref{14}), the two-iteration approximation (\ref%
{10}) or the $n-$independent approximation (\ref{16})?

Also, it is interesting to compare these variants of Garrett's approximation
with another simple result for the energy of the bound state in a finite
rectangular well - Barker's formula. Let us remind that these two
approximations are obtained from two different perspectives: Garrett
proposes a physical idea (the existence of a penetration depth); Barker et
al. use a mathematical approximation (transforming the transcendental
eigenvalue equations into easily solvable, low order algebraic equations).

In order to analyze these issues, we shall calculate the errors generated by
each variant. Let us firstly introduce convenient notations. The energy of a
bound state, in any variant of Garrett's approximation is, according to (\ref%
{4}) or (\ref{6}):

\begin{equation}
E_{n}=\frac{\hbar ^{2}k_{n}^{2}}{2m}=n^{2}\frac{\pi ^{2}\hbar ^{2}}{2m\left(
L+2\delta _{n}\right) ^{2}}=\frac{\pi ^{2}\hbar ^{2}n^{2}}{2mL^{2}}\frac{1}{%
\left( 1+y_{n}\right) ^{2}},\ y_{n}=\frac{2\delta _{n}}{L}  \label{17}
\end{equation}%
and it can be expressed in terms of the dimensionless wave vector

\begin{equation}
K_{n}=Lk_{n}=\frac{\pi n}{1+y_{n}}  \label{18}
\end{equation}%
We have to distinguish among three different formulas for $K$, corresponding
to each of the three variants of Garrett's approximation:

(1) the two iterations Garrett approximation, used in his original paper,

\begin{equation}
K^{\left( 2\right) }\left( P,n\right) =\frac{\pi n}{1+y^{\left( 2\right)
}\left( P,n\right) },\ y^{\left( 2\right) }\left( P,n\right) =\frac{2\delta
_{n}^{\left( 2\right) }}{L}  \label{19}
\end{equation}%
where $\delta _{n}^{\left( 2\right) }$\ is defined in (\ref{10});

(2) the "consistent" Garrett approximation, obtained after infinitely many
iterations:

\begin{equation}
K_{4}\left( P,n\right) =\frac{\pi n}{1+y_{4}\left( P,n\right) }  \label{20}
\end{equation}%
where $y_{4}\left( P,n\right) $ is the root of the quartic equation (\ref{14}%
), and

(3) the lowest order Garrett approximation, given by the root (\ref{16}) of
eq. (\ref{15}):

\begin{equation}
K_{0}\left( P,n\right) =\frac{\pi n}{1+y_{0}\left( P\right) }  \label{21}
\end{equation}

In order to compare the various Garrett approximations with Barker formula,
we shall also define:%
\begin{equation}
K_{B}\left( P,n\right) =2\alpha _{n}=\frac{2P}{1+P}\left( \frac{n\pi }{2}-%
\frac{1}{6\left( 1+P\right) ^{3}}\left( \frac{n\pi }{2}\right) ^{3}\right)
\label{22}
\end{equation}%
where $\alpha _{n}$ refers to Barker's notation, eq. (\ref{16}) of \cite%
{[Barker]}.

The exact value of the dimensionless wave vector, i.e. the solution of the
equation%
\begin{equation}
\frac{K}{2}=\frac{n\pi }{2}-\arcsin \frac{K}{2P}\ \   \label{23}
\end{equation}%
will be denoted $K_{ex}\left( P,n\right) .$ The errors of the aforementioned
approximations are defined as:

\begin{equation*}
\varepsilon ^{\left( 2\right) }\left( P,n\right) =\frac{K_{ex}\left(
P,n\right) -K^{\left( 2\right) }\left( P,n\right) }{K_{ex}\left( P,n\right) }
\end{equation*}

\begin{equation}  \label{24}
\end{equation}

\begin{equation*}
\varepsilon _{a}\left( P,n\right) =\frac{K_{ex}\left( P,n\right)
-K_{a}\left( P,n\right) }{K_{ex}\left( P,n\right) },\ with\ a=4,\ 0,\ B
\end{equation*}

\section{Comparative analysis of Garrett and Barker approximations for
finite square wells}

The numerical values of the errors of the three variants of Garrett
approximation and of the dimensionless characteristic (penetration)
lengths $y^{\left( 2\right) }\left( P,n\right) ,\ \ y_{4}\left( P,n\right)$
 for $P=1,...,10$ and for any $n$ characterizing each bound state, are
given as auxiliary material. For a well with $P=10$, the plots of the
absolute values of errors $\varepsilon _{4},\varepsilon _{0}$ for $%
n=1,2,...7 $ and of $\varepsilon ^{\left( 2\right) }$ for $n=1,2,...6$
(this approximation is unphysical for $n=7$) are given in
Fig. 1. Any other similar plot can be easily done, using the auxiliary material or
the formulas (\ref{19}-\ref{21}).

The conclusions of this analysis are quite surprising. The consistent
Garrett approximation is really useful only for shallow wells $\left(
P=1\right) ,$ where it is much better even than Barker's one, and the two
iteration approximation is unphysical (complex). Otherwise, it is less
precise then (or comparable to) the two-iteration approximation; actually,
the main inconvenient of the two-iteration approach is that it is unphysical
(complex) for the highest level of any of the wells examined here. Even more
surprising is that the lowest order approximation is the most precise one
(among the Garrett approximations), for highest levels; let us remind that
it was obtained using approximations valid for deep wells (large $P$) and
deep levels (small $n$). Typically, the lowest levels are better described
by the two iteration approximation; the few exceptions, when the
"consistent" approximation is more precise, are numerically irrelevant, for
instance: $\varepsilon _{4}\left( P=7,n=1\right) =1.6488\times
10^{-4}<\varepsilon ^{\left( 2\right) }\left( P=7,n=1\right) =1.6747\times
10^{-4}.$ Actually, excepting the case of shallow wells $\left( P=1\right) ,$
the only benefit of the consistent Garrett approximation is that it
generates the lowest order approximation, which is surprisingly accurate!

To conclude, the responses to the questions put in the previous section are
the following:

(*1) the consistent approximation is the only one to give good results for
shallow wells $\left( P\sim 1\right) ,$ and Garrett's original two-iteration
approach is the most accurate for relatively low levels $\left( n\lesssim
\frac{n_{\max }}{2}\right) $; actually, it is unphysical for the highest
level $\left( n\sim n_{\max }\right) $

(*2) the $n-$independent approximation is the most accurate one for
relatively high levels $\left( \frac{n_{\max }}{2}\lesssim n\lesssim n_{\max
}\right) $

(*3) if we are interested in accuracy, we should made a case by case
analysis, eventually guided by the additional materials, as
there is no general rule; if we are interested in the simplest analytical
approximation, we should choose the $n-$independent approximation.

One more remark: taking into account the validity of mathematical
approximations done in order to obtain Barker approximation, it is supposed
to work well for large $P$ and relatively small $n$; actually, as we can see
from Tables 1-4, it gives excellent results for $P=2$ and
for any larger $P$, if $n$ is relatively high.

\section{The simple asymmetric well}

Let as consider the simplest generalization of the symmetric rectangular
well, called sometimes simple asymmetric square well. Its
corresponding Schroedinger equation:

\begin{equation}
\left( -\frac{\hbar ^{2}}{2m}\frac{d^{2}}{dx^{2}}+V\left( x\right) \right)
\psi =E\psi  \label{25}
\end{equation}%
can be written simpler, as

\begin{equation}
\psi ^{\prime \prime }+\left[ k^{2}-U\left( x\right) \right] \psi =0
\label{26}
\end{equation}%
if we introduce $U\left( x\right) $ instead of $V\left( x\right) $\ by:

\begin{equation}
V\left( x\right) =\frac{\hbar ^{2}}{2m}U\left( x\right) .  \label{27}
\end{equation}%
We shall define, following Messiah \cite{[Messiah]}, Ch. III, \S 6 (see also
\cite{[LL]}, \S 22, problem 2)

\begin{equation}
U\left( x\right) =U_{3}\theta \left( b-x\right) +U_{2}\theta \left(
x-b\right) \theta \left( a-x\right) +U_{1}\theta \left( x-2\right)
\label{28}
\end{equation}%
where $\theta $ is the Heaveside function.

The bound state wave function has the form:

\begin{equation}
\psi \left( x\right) =\left\{
\begin{array}{c}
A_{1}e^{-K_{1}x},\ x>a \\
A_{2}\sin \left( kx+\varphi \right) ,\ b<a<a \\
A_{3}e^{K_{3}x},\ x<b%
\end{array}%
\right.  \label{29}
\end{equation}

\bigskip We shall put:

\begin{equation}
K_{2}=\sqrt{k^{2}-U_{2}},\ K_{1}=\sqrt{U_{1}-k^{2}},\ K_{3}=\sqrt{U_{3}-k^{2}%
}  \label{30}
\end{equation}

Without restricting the generality, we can choose $U_{2}=0$ and define:

\begin{equation}
L=b-a,\ P_{1}=\sqrt{2mU_{1}}\frac{L}{2\hbar },\ \ P_{3}=\sqrt{2mU_{3}}\frac{L%
}{2\hbar }  \label{31}
\end{equation}

The eigenvalue equation associated to the solution (\ref{29}) has the form:

\begin{equation}
n\pi -Lk=\arcsin \frac{Lk}{2P_{3}}+\arcsin \frac{Lk}{2P_{1}}  \label{32}
\end{equation}%
For a symmetric well, $\ P_{1}=P_{3}$ and (\ref{32}) becomes:

\begin{equation}
n\pi -Lk=2\arcsin \frac{Lk}{2P}  \label{33}
\end{equation}

identical with (\ref{23}).

As Garrett noticed, the approach used for the symmetric wells can be also
applied here, for the $n-$th bound state, with $L$ replaced by $L+\delta
_{l}+\delta _{r}=L\left( 1+y_{l}+y_{r}\right) $ (where the indices $l$ and $%
r $ refer to the left, respectively right wall). The dimensionless
penetration depths $y_{l},\ y_{r}$ are evaluated choosing that variant of
Garrett approximation with smallest error, for a given pair $\left(
P,n\right) ,$ characterizing the $n-$th bound state of a square well of
strength $P.$

To see how the method works, let us consider the case $P_{3}=10,\ P_{1}=8,$
with $n_{\max }=6$ bound states. For $\left( P_{3}=10,\ n=1\right) ,$ the
most accurate variant of Garrett approximation for a rectangular well gives $%
y_{4}\left( P_{3}=10,n=1\right) =0.10103,$ and for $\left( P_{1}=8,\
n=1\right) $, the best one is $y_{4}\left( P_{1}=8,n=1\right) =0.126942,$ so
the Garrett approximation for the asymmetric well gives the dimensionless
wave vector:

\begin{equation*}
K_{ap}\left( P_{3}=10,P_{1}=8;\ n=1\right) =
\end{equation*}%
\begin{equation}  \label{34}
\end{equation}%
\begin{equation*}
=\frac{\pi }{1+\frac{1}{2}\left( y_{4}\left( P_{3}=10,n=1\right)
+y_{4}\left( P_{1}=8,n=1\right) \right) }=2.\,\allowbreak 820\,1
\end{equation*}%
to be compared to the "exact" value of (\ref{32}) with the same parameters,

$K_{ex}\left( P_{3}=10,P_{1}=8;\ n=1\right) =2.82264.$ For $%
P_{3}=10,P_{1}=8;\ n=2,$ the most accurate variant is again, for both walls,
$y_{4},$ but for $P_{3}=10,P_{1}=8;\ n=3,$ for the both walls, the most
accurate variant is the $n-$independent one, so $%
y_{0}(P_{3}=10,n=3)=y_{0}(P_{3}=10)$ and $y_{0}(P_{1}=8,n=3)=y_{0}(P_{1}=8).$
The approximate value is:

\begin{equation*}
K_{ap}\left( P_{3}=10,P_{1}=8;\ n=3\right) =
\end{equation*}%
\begin{equation}  \label{35}
\end{equation}%
\begin{equation*}
=\frac{3\pi }{1+\frac{1}{2}\left( y_{0}\left( P_{3}=10\right) +y_{0}\left(
P_{1}=8\right) \right) }=8.\,\allowbreak 373\,6
\end{equation*}%
to be compared to the "exact" one, $K_{ex}\left( P_{3}=10,P_{1}=8;\
n=3\right) =8.4342.$ Actually, the errors are, in these two cases:

\begin{equation}
\varepsilon _{\left( P_{3}=10,P_{1}=8;\ n=1\right) }=\left( \frac{%
K_{ex}-K_{ap}}{K_{ex}}\right) _{\left( P_{3}=10,P_{1}=8;\ n=1\right)
}=8.\,\allowbreak 998\,7\times 10^{-4}  \label{36}
\end{equation}%
(let us mention that $\varepsilon _{0}\left( P_{3}=10,\ n=3\right)
=6.02\times 10^{-3}\ $and $\varepsilon _{0}\left( P_{3}=8,\ n=3\right)
=8.02\times 10^{-4}$) and

\begin{equation}
\varepsilon _{\left( P_{3}=10,P_{1}=8;\ n=3\right) }=\left( \frac{%
K_{ex}-K_{ap}}{K_{ex}}\right) _{\left( P_{3}=10,P_{1}=8;\ n=3\right)
}=7.\,\allowbreak 1850\times 10^{-3}  \label{37}
\end{equation}%
(let us mention that $\varepsilon _{4}\left( P_{3}=10,\ n=3\right)
=6.27\times 10^{-4}\ $and $\varepsilon _{4}\left( P_{3}=8,\ n=3\right)
=1.15\times 10^{-4}$). So, the error of the $n-$th bound state energy in the
asymmetric well with strengths $\left( P_{3},P_{1}\right) $ is comparable to
the error of the most precise variant of the Garrett approximation of the $%
n- $th bound state of the symmetric wells with strength $P_{3},\ P_{1}.$ The
correctness of this empirical remark was verified in all cases we worked out
(see the auxiliary material).

The errors for other values of $n$ are plotted in Fig.1, and can be easily obtained,
for any pair $\left( P_{3},P_{1}\right) $, using the
auxiliary material.
\begin{figure}[tbp]
\begin{center}
\includegraphics[width=\textwidth]{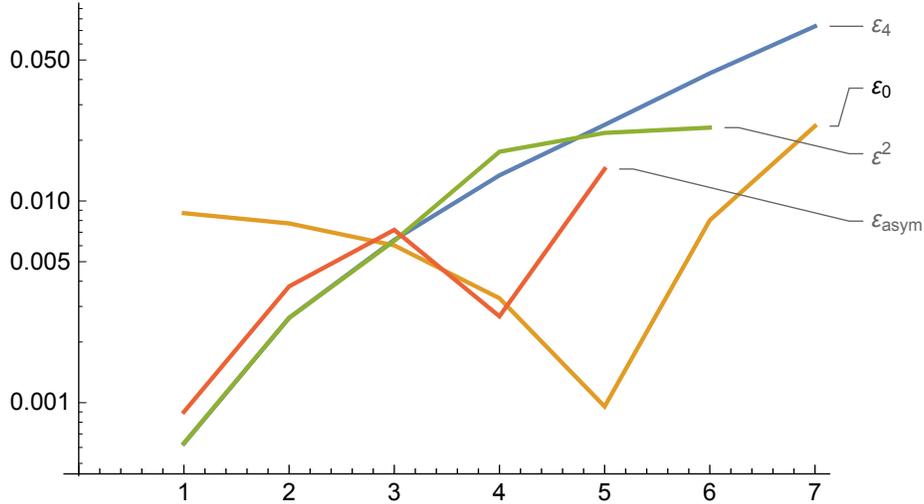}
\end{center}
\caption{ The errors $\varepsilon _{4}$, $\varepsilon _{0}$, $\varepsilon ^{(2)}$ of the n-th bound state energy for a square well with P=10, as functions of n, and, similarly, the error $\varepsilon _{asym}$ for the bound states an asymmetric well with $P_{3}=10,P_{1}=8$.}
\end{figure}

\section{Applications}

We shall shortly describe here some of the applications mentioned in the
Introduction.

Garrett's idea was used in replacing "a stepped spherical potential" with an
effective, impenetrable one, in order to calculate the thermodynamic
properties of a system of non-interacting bosons confined in a quantum dot.
The same approach was discussed in the context of semiconductor quantum dots
\cite{[Ind-1999]}.

In a study of interference effects in capillary neutron guides \cite%
{[Roh-2002]}, Rohwedder examines both circular and rectangular cases. In the
circular (cylindrical) case, the neutrons effectively "see" a reflecting
wall not at the radius $R,$ but a slightly larger "effective" radius $%
R\,_{eff}\simeq R+d$. This can be interpreted as a waveguide-confined
manifestation of the Goos-H\"{a}nchen effect \cite{[Goos-Han]}. For
rectangular guides, with section $\left( a_{x},a_{y}\right) ,$ the variables
can be easily separated, and the energy eigenvalues are approximately given
by the corresponding spectrum of an infinite square well; its "effective"
width $a_{x,eff}\simeq a_{x}+d,\ a_{y,eff}\simeq a_{y}+d$ turns out to be
slightly larger than the "bare" width. The amount $d=\hbar /\sqrt{2MV}$ can
once more be identified with the evanescent penetration depth of the
lowest-lying eigenmodes, and - again - is an expression of the
(waveguide-confined) Goos-H\"{a}nchen effect.

Finally, we can expect that, applying to a stepped rectangular well, an
approach similar to that used in the previous section for asymmetric wells
(i.e. associating to each wall a strength $P$ and a penetration depth $%
\delta $), we shall obtain similar accuracy in the evaluation of bound
states energy.

\section{Conclusions}

Essentially, Garrett approximations consists in the following steps: (1) for
the $n-$th bound state of a particle in a rectangular well, and for each
wall of the well, we associate a penetration depth; (2) in this way, we
define a larger, "effective" well, with impenetrable walls; (3) the $n-$th
level of this (infinite) well is a good approximation for the $n-$th level
of the finite well.

We discussed in detail the three variants of this approximation and calculated
its errors in a large number of cases; the error of Barker approximation,
one of the most precise alternative approximations, is also obtained - it is
typically smaller than Garrett's. Quite surprisingly, the simplest variants give the most
accurate results. The method works almost equally well for symmetric and
asymmetric wells (slightly better, in the symmetric case). The applications
for quantum wells, quantum dots and capillary neutron wave guides are
shortly discussed.

Garrett approximation is an analytical one, based on a simple physical idea,
and this is why it can be extended to more complicated rectangular
potentials. One could object that it is unnecessary to use such an
approximation, when a very precise result can be easily obtained
numerically, but an analytic formula remains attractive, especially in
this case, when its form - based on a result obtained for infinite wells -
is so simple.

\newpage
\section{Auxiliary material}

{Table 1. $\left( P=1-5\right) $}
\vskip0.4cm
\begin{tabular}{||c||c||c||c||c||c||}
\hline\hline
$\varepsilon _{4}$ & $\varepsilon _{0}$ & $\varepsilon ^{\left( 2\right) }$
& $y_{4}$ & $y^{\left( 2\right) }$ & $\varepsilon _{B}$ \\ \hline\hline
\multicolumn{6}{||c||}{$P=1,n=1$} \\ \hline\hline
$9.\,\allowbreak 415\,4\times 10^{-2}$ & $0.291\,56$ & $-$ & $1.3462$ & $-$
& $-0.112\,85$ \\ \hline\hline
\multicolumn{6}{||c||}{$P=2,n=1$} \\ \hline\hline
$3.\,\allowbreak 259\,9\times 10^{-2}$ & $0.128\,43$ & $1.\,\allowbreak
922\,6\times 10^{-2}$ & $0.5766$ & $0.555\,13$ & $-1.\,\allowbreak
359\,4\times 10^{-3}$ \\ \hline\hline
\multicolumn{6}{||c||}{$P=2,n=2$} \\ \hline\hline
$0.125\,12$ & $5.\,\allowbreak 291\,2\times 10^{-2}$ & $-$ & $0.8944$ & $-$
& $-3.\,\allowbreak 761\,5\times 10^{-2}$ \\ \hline\hline
\multicolumn{6}{||c||}{$P=3,n=1$} \\ \hline\hline
$1.\,\allowbreak 373\,8\times 10^{-2}$ & $7.\,\allowbreak 065\,2\times
10^{-2}$ & $1.\,\allowbreak 273\,4\times 10^{-2}$ & $0.3611$ & $0.4367$ & $%
-3.\,\allowbreak 845\,8\times 10^{-4}$ \\ \hline\hline
\multicolumn{6}{||c||}{$P=3,n=2$} \\ \hline\hline
$0.210\,24$ & $4.\,\allowbreak 559\,8\times 10^{-2}$ & $-$ & $0.7456$ & $-$
& $-7.\,\allowbreak 372\,1\times 10^{-3}$ \\ \hline\hline
\multicolumn{6}{||c||}{$P=4,n=1$} \\ \hline\hline
$6.\,\allowbreak 936\,5\times 10^{-3}$ & $4.\,\allowbreak 641\,8\times
10^{-2}$ & $6.\,\allowbreak 787\,2\times 10^{-3}$ & $0.2630$ & $0.262\,84$ &
$-1.\,\allowbreak 197\,7\times 10^{-4}$ \\ \hline\hline
\multicolumn{6}{||c||}{$P=4,n=2$} \\ \hline\hline
$3.\,\allowbreak 242\,2\times 10^{-2}$ & $3.\,\allowbreak 272\,3\times
10^{-2}$ & $2.\,\allowbreak 463\,1\times 10^{-2}$ & $0.3121$ & $0.301\,62$ &
$-2.\,\allowbreak 283\,2\times 10^{-3}$ \\ \hline\hline
\multicolumn{6}{||c||}{$P=4,n=3$} \\ \hline\hline
$9.\,\allowbreak 616\,8\times 10^{-2}$ & $1.\,\allowbreak 366\,1\times
10^{-3}$ & $-$ & $0.4367$ & $-$ & $-1.\,\allowbreak 752\,3\times 10^{-2}$ \\
\hline\hline
\multicolumn{6}{||c||}{$P=5,n=1$} \\ \hline\hline
$3.\,\allowbreak 967\,4\times 10^{-3}$ & $0.030\,38$ & $3.\,\allowbreak
942\,0\times 10^{-3}$ & $0.2071$ & $0.207\,09$ & $-3.\,\allowbreak
827\,2\times 10^{-5}$ \\ \hline\hline
\multicolumn{6}{||c||}{$P=5,n=2$} \\ \hline\hline
$1.\,\allowbreak 800\,5\times 10^{-2}$ & $2.\,\allowbreak 395\,8\times
10^{-2}$ & $1.\,\allowbreak 673\,9\times 10^{-2}$ & $0.2325$ & $0.230\,91$ &
$-9.\,\allowbreak 053\,4\times 10^{-4}$ \\ \hline\hline
\multicolumn{6}{||c||}{$P=5,n=3$} \\ \hline\hline
$4.\,\allowbreak 977\,0\times 10^{-2}$ & $9.\,\allowbreak 685\,3\times
10^{-3}$ & $1.\,\allowbreak 574\,0\times 10^{-2}$ & $0.2923$ & $0.247\,63$ &
$-5.\,\allowbreak 798\,1\times 10^{-3}$ \\ \hline\hline
\multicolumn{6}{||c||}{$P=5,n=4$} \\ \hline\hline
$0.101\,05$ & $-3.\,\allowbreak 275\,5\times 10^{-2}$ & $-$ & $0.4246$ & $-$
& $-3.\,\allowbreak 470\,1\times 10^{-2}$ \\ \hline\hline
\end{tabular}

\newpage
{Table 2. $\left( P=6,7\right) $}
\vskip0.4cm
\begin{tabular}{||c||c||c||c||c||c||}
\hline\hline
$\varepsilon _{4}$ & $\varepsilon _{0}$ & $\varepsilon ^{\left( 2\right) }$
& $y_{4}$ & $y^{\left( 2\right) }$ & $\varepsilon _{B}$ \\ \hline\hline
\multicolumn{6}{||c||}{$P=6,n=1$} \\ \hline\hline
$3.\,\allowbreak 468\,9\times 10^{-2}$ & $2.\,\allowbreak 204\,9\times
10^{-2}$ & $2.\,\allowbreak 197\,4\times 10^{-2}$ & $0.1710$ & $0.170\,98$ &
$-3.\,\allowbreak 718\,2\times 10^{-5}$ \\ \hline\hline
\multicolumn{6}{||c||}{$P=6,n=2$} \\ \hline\hline
$1.\,\allowbreak 095\,4\times 10^{-2}$ & $1.\,\allowbreak 815\,5\times
10^{-2}$ & $1.\,\allowbreak 063\,9\times 10^{-2}$ & $0.1858$ & $0.185\,39$ &
$-4.\,\allowbreak 293\times 10^{-4}$ \\ \hline\hline
\multicolumn{6}{||c||}{$P=6,n=3$} \\ \hline\hline
$2.\,\allowbreak 936\,1\times 10^{-2}$ & $0.010\,18$ & $2.\,\allowbreak
465\,0\times 10^{-2}$ & $0.2180$ & $0.212\,16$ & $-2.\,\allowbreak
446\,2\times 10^{-3}$ \\ \hline\hline
\multicolumn{6}{||c||}{$P=6,n=4$} \\ \hline\hline
$0.065\,66$ & $-6.\,\allowbreak 608\,6\times 10^{-3}$ & $-$ & $0.2868$ & $-$
& $-1.\,\allowbreak 081\,9\times 10^{-2}$ \\ \hline\hline
\multicolumn{6}{||c||}{$P=7,n=1$} \\ \hline\hline
$1.\,\allowbreak 648\,8\times 10^{-3}$ & $0.016\,73$ & $1.\,\allowbreak
674\,7\times 10^{-3}$ & $0.1457$ & $0.145\,67$ & $-1.\,\allowbreak
225\,1\times 10^{-5}$ \\ \hline\hline
\multicolumn{6}{||c||}{$P=7,n=2$} \\ \hline\hline
$7.\,\allowbreak 150\,1\times 10^{-3}$ & $1.\,\allowbreak 417\,7\times
10^{-2}$ & $7.\,\allowbreak 045\,2\times 10^{-3}$ & $0.1550$ & $0.154\,93$ &
$-2.\,\allowbreak 190\,2\times 10^{-4}$ \\ \hline\hline
\multicolumn{6}{||c||}{$P=7,n=3$} \\ \hline\hline
$1.\,\allowbreak 855\,4\times 10^{-2}$ & $9.\,\allowbreak 203\,7\times
10^{-3}$ & $1.\,\allowbreak 745\,1\times 10^{-2}$ & $0.1743$ & $0.173\,03$ &
$-1.\,\allowbreak 198\,4\times 10^{-3}$ \\ \hline\hline
\multicolumn{6}{||c||}{$P=7,n=4$} \\ \hline\hline
$4.\,\allowbreak 066\,9\times 10^{-2}$ & $7.\,\allowbreak 070\,1\times
10^{-5}$ & $2.\,\allowbreak 604\,7\times 10^{-2}$ & $0.2125$ & $0.194\,32$ &
$-4.\,\allowbreak 683\,7\times 10^{-3}$ \\ \hline\hline
\multicolumn{6}{||c||}{$P=7,n=5$} \\ \hline\hline
$7.\,\allowbreak 953\,7\times 10^{-2}$ & $-2.\,\allowbreak 051\,1\times
10^{-2}$ & $-$ & $0.2897$ & $-$ & $-1.\,\allowbreak 784\,3\times 10^{-2}$ \\
\hline\hline
\end{tabular}

\newpage
{Table 3. $\left( P=8,9\right) $}
\vskip0.4cm
\begin{tabular}{||c||c||c||c||c||c||}
\hline\hline
$\varepsilon _{4}$ & $\varepsilon _{0}$ & $\varepsilon ^{\left( 2\right) }$
& $y_{4}$ & $y^{\left( 2\right) }$ & $\varepsilon _{B}$ \\ \hline\hline
\multicolumn{6}{||c||}{$P=8,n=1$} \\ \hline\hline
$1.\,\allowbreak 152\,6\times 10^{-3}$ & $1.\,\allowbreak 313\,5\times
10^{-2}$ & $1.\,\allowbreak 146\,6\times 10^{-3}$ & $0.1269$ & $0.126\,94$ &
$-3.\,\allowbreak 583\,1\times 10^{-5}$ \\ \hline\hline
\multicolumn{6}{||c||}{$P=8,n=2$} \\ \hline\hline
$4.\,\allowbreak 924\,9\times 10^{-3}$ & $1.\,\allowbreak 135\,3\times
10^{-2}$ & $4.\,\allowbreak 881\,7\times 10^{-3}$ & $0.13325$ & $0.133\,21$
& $-1.\,\allowbreak 256\,3\times 10^{-4}$ \\ \hline\hline
\multicolumn{6}{||c||}{$P=8,n=3$} \\ \hline\hline
$1.\,\allowbreak 245\,0\times 10^{-2}$ & $8.\,\allowbreak 024\,7\times
10^{-3}$ & $1.\,\allowbreak 210\,1\times 10^{-2}$ & $0.1457$ & $0.14533$ & $%
-6.\,\allowbreak 362\,8\times 10^{-4}$ \\ \hline\hline
\multicolumn{6}{||c||}{$P=8,n=4$} \\ \hline\hline
$2.\,\allowbreak 637\,6\times 10^{-2}$ & $2.\,\allowbreak 336\,4\times
10^{-3}$ & $2.\,\allowbreak 335\,4\times 10^{-2}$ & $0.1688$ & $0.165\,13$ &
$-2.\,\allowbreak 363\,5\times 10^{-3}$ \\ \hline\hline
\multicolumn{6}{||c||}{$P=8,n=5$} \\ \hline\hline
$5.\,\allowbreak 199\,5\times 10^{-2}$ & $-8.\,\allowbreak 053\,1\times
10^{-3}$ & $4.\,\allowbreak 633\,5\times 10^{-3}$ & $0.2129$ & $0.155\,16$ &
$-7.\,\allowbreak 664\,0\times 10^{-3}$ \\ \hline\hline
\multicolumn{6}{||c||}{$P=8,n=6$} \\ \hline\hline
$0.239\,53$ & $0.239\,53$ & $-$ & $0.29789$ & $-$ & $-3.\,\allowbreak
144\,3\times 10^{-2}$ \\ \hline\hline
\multicolumn{6}{||c||}{$P=9,n=1$} \\ \hline\hline
$8.\,\allowbreak 372\,4\times 10^{-4}$ & $1.\,\allowbreak 057\,8\times
10^{-2}$ & $8.\,\allowbreak 138\,1\times 10^{-4}$ & $0.112\,5$ & $0.112\,5$
& $-3.\,\allowbreak 538\,3\times 10^{-5}$ \\ \hline\hline
\multicolumn{6}{||c||}{$P=9,n=2$} \\ \hline\hline
$3.\,\allowbreak 537\,6\times 10^{-3}$ & $9.\,\allowbreak 291\,5\times
10^{-3}$ & $3.\,\allowbreak 525\,1\times 10^{-3}$ & $0.116\,70$ & $0.116\,95$
& $-7.\,\allowbreak 085\,7\times 10^{-5}$ \\ \hline\hline
\multicolumn{6}{||c||}{$P=9,n=3$} \\ \hline\hline
$8.\,\allowbreak 763\,7\times 10^{-3}$ & $6.\,\allowbreak 942\,8\times
10^{-3}$ & $8.\,\allowbreak 629\,6\times 10^{-3}$ & $0.1345$ & $0.125\,4$ & $%
-3.\,\allowbreak 788\times 10^{-4}$ \\ \hline\hline
\multicolumn{6}{||c||}{$P=9,n=4$} \\ \hline\hline
$2.\,\allowbreak 236\,8\times 10^{-3}$ & $-1.\,\allowbreak 290\,9\times
10^{-2}$ & $1.\,\allowbreak 349\,3\times 10^{-3}$ & $0.1405$ & $0.139\,46$ &
$-1.\,\allowbreak 739\,6\times 10^{-2}$ \\ \hline\hline
\multicolumn{6}{||c||}{$P=9,n=5$} \\ \hline\hline
$3.\,\allowbreak 445\,1\times 10^{-2}$ & $-3.\,\allowbreak 291\,7\times
10^{-3}$ & $2.\,\allowbreak 67\times 10^{-2}$ & $0.1673$ & $0.157\,99$ & $%
-3.\,\allowbreak 960\,7\times 10^{-3}$ \\ \hline\hline
\multicolumn{6}{||c||}{$P=9,n=6$} \\ \hline\hline
$6.\,\allowbreak 319\,5\times 10^{-2}$ & $-1.\,\allowbreak 540\,4\times
10^{-2}$ & $-$ & $0.2177$ & $-$ & $-1.\,\allowbreak 146\,2\times 10^{-2}$ \\
\hline\hline
&  &  &  &  &  \\ \hline\hline
\end{tabular}

\newpage
{Table 4. $\left( P=10\right) $}
\vskip0.4cm
\begin{tabular}{||c||c||c||c||c||c||}
\hline\hline
$\varepsilon _{4}$ & $\varepsilon _{0}$ & $\varepsilon ^{\left( 2\right) }$
& $y_{4}$ & $y^{\left( 2\right) }$ & $\varepsilon _{B}$ \\ \hline\hline
\multicolumn{6}{||c||}{$P=10,n=1$} \\ \hline\hline
$6.\,\allowbreak 273\,0\times 10^{-4}$ & $8.\,\allowbreak 700\,2\times
10^{-3}$ & $6.\,\allowbreak 304\,5\times 10^{-4}$ & $0.101\,03$ & $0.101\,03$
& $-2.\,\allowbreak 577\,7\times 10^{-6}$ \\ \hline\hline
\multicolumn{6}{||c||}{$P=10,n=2$} \\ \hline\hline
$2.\,\allowbreak 627\,8\times 10^{-3}$ & $7.\,\allowbreak 740\,4\times
10^{-3}$ & $2.\,\allowbreak 629\,4\times 10^{-3}$ & $0.104\,3$ & $0.104\,3$
& $-3.\,\allowbreak 505\,9\times 10^{-5}$ \\ \hline\hline
\multicolumn{6}{||c||}{$P=10,n=3$} \\ \hline\hline
$6.\,\allowbreak 409\,0\times 10^{-3}$ & $6.\,\allowbreak 017\,2\times
10^{-3}$ & $6.\,\allowbreak 345\,0\times 10^{-3}$ & $0.1104$ & $0.110\,37$ &
$-2.\,\allowbreak 341\,3\times 10^{-4}$ \\ \hline\hline
\multicolumn{6}{||c||}{$P=10,n=4$} \\ \hline\hline
$0.012\,86$ & $3.\,\allowbreak 289\,3\times 10^{-3}$ & $1.\,\allowbreak
745\,1\times 10^{-2}$ & $0.1208$ & $0.120\,38$ & $-7.\,\allowbreak
571\,5\times 10^{-4}$ \\ \hline\hline
\multicolumn{6}{||c||}{$P=10,n=5$} \\ \hline\hline
$2.\,\allowbreak 381\,5\times 10^{-2}$ & $-9.\,\allowbreak 574\,2\times
10^{-4}$ & $2.\,\allowbreak 165\,8\times 10^{-2}$ & $0.1382$ & $0.135\,73$ &
$-2.\,\allowbreak 277\,6\times 10^{-3}$ \\ \hline\hline
\multicolumn{6}{||c||}{$P=10,n=6$} \\ \hline\hline
$4.\,\allowbreak 287\,1\times 10^{-2}$ & $-8.\,\allowbreak 023\,9\times
10^{-3}$ & $2.\,\allowbreak 305\,5\times 10^{-2}$ & $0.1690$ & $0.145\,29$ &
$-5.\,\allowbreak 852\,9\times 10^{-3}$ \\ \hline\hline
\multicolumn{6}{||c||}{$P=10,n=7$} \\ \hline\hline
$7.\,\allowbreak 341\,7\times 10^{-2}$ & $-2.\,\allowbreak 345\,7\times
10^{-2}$ & $-$ & $0.2261$ & $-$ & $-1.\,\allowbreak 712\,6\times 10^{-2}$ \\
\hline\hline
\end{tabular}

\bigskip

\begin{acknowledgement}
The author acknowledges the financial support of the ANCSI - IFIN-HH project
PN 18 09 01 01/2018.
\end{acknowledgement}

\end{document}